\documentclass{elsart4}

\usepackage{graphics}
\usepackage{graphicx}
\usepackage{epsfig}
\usepackage{amssymb}

\begin{document}

\begin{frontmatter}

\title{Measurement of mesoscopic High-$T_c$ superconductors using Si mechanical micro-oscillators}

\author{M. Dolz , D. Antonio, H. Pastoriza}
\address{Centro At\'omico Bariloche, Comisi\'on Nacional de Energ\'{\i}a At\'omica and CONICET, R8402AGP S.
C. de Bariloche, Argentina}

\begin{abstract}

In a superconducting mesoscopic sample, with dimensions comparable
to the London penetration depth, some properties are qualitatively
different to those found in the bulk material. These properties
include magnetization, vortex dynamics and ordering of the vortex
lattice. In order to detect the small signals produced by this
kind of samples, new instruments designed for the microscale are
needed. In this work we use micromechanical oscillators to study
the magnetic properties of a Bi$_2$Sr$_2$CaCu$_2$O$_{8 + \delta}$
disk with a diameter of $13.5$ microns and a thickness of $2.5$
microns. The discussion of our results is based on the existence
and contribution of inter and intra layer currents.
\end{abstract}

\end{frontmatter}

\section{Introduction}
\label{}

The study of the vortex physics in mesoscopic samples \cite{Aksyuk98,Chan01} is difficult
due to the need of instruments sensible enough to detect their signals. Sensitive
instruments such as SQUIDs are not the best option because they are not designed to
measure microscopic samples. The use of mechanical oscillators as magnetometers is not a
new idea, they have been used successfully for this application for some time
\cite{Kleiman85}. Our approach for studying mesoscopic high $T_c$ samples is to use
silicon micro-oscillators (following the work of \cite{Bolle99}) which have a torsional
mode with a resonant frequency $\nu _r \approx 45$ kHz and a quality factor $Q > 10^4$ at
low temperatures. This instrument integrates high sensitivity and reduced size with a
small signal loss, which is an important factor in the measurement of micron sized
samples.
\begin{figure}[t]\begin{center}
\includegraphics[width=\linewidth]{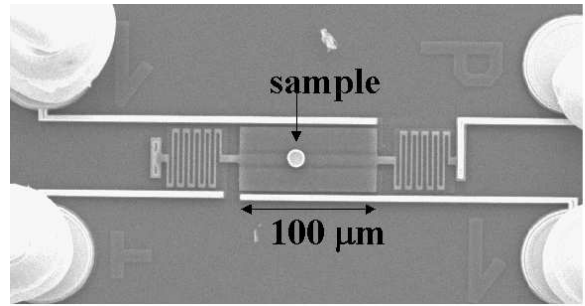}
\caption{Scanning electron micrograph of a high-$Q$ mechanical
oscillator with a BSCCO disk with a diameter of $13.5$ microns and
a thickness of $2.5$ microns. The sample was glued by means of
micro-pipettes and micro-manipulators.} \label{oscillator}
\end{center}\end{figure}

In this work we present the response of a micro-oscillator with a
Bi$_2$Sr$_2$CaCu$_2$O$_{8 + \delta}$ (BSCCO) disk mounted on top
of it, in the presence of an external magnetic field $H$. The
system (oscillator and sample) is shown in Fig. \ref{oscillator}.

\section{Experimental details}
\label{}

\begin{figure}[t]\begin{center}
\includegraphics[width=\linewidth]{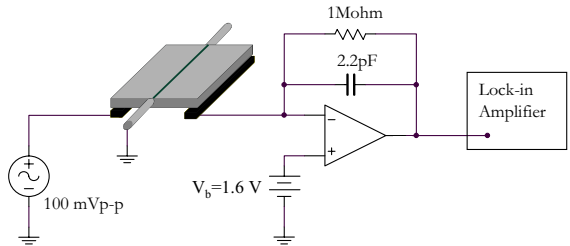}
\caption{System under study. The plate is driven by a sinusoidal
voltage. The displacement is capacitively detected with a
Current-Voltage Converter and a Lock-in Amplifier.}
\label{circuit}
\end{center}\end{figure}

The mechanical micro-oscillator was manufactured at MEMSCAP \cite{Memscap} following the
MUMPS specifications. It consists of a central plate connected to two serpentine springs
which are anchored to the substrate. Below the plate two separate electrodes carry the
electrical signals. The plate is electrically grounded and harmonically driven by one of
the electrodes. The other electrode is used to detect the amplitude and phase of the
mechanical oscillations capacitively. The plate and the detection electrode form a
capacitor of $\approx 10$ fF. The motion of the plate produces a variation in the
capacitance $\delta C < 1$ fF. A bias voltage $V_b = 1.6$ V is held constant in the
capacitor and the current, proportional to $\delta C$ is measured by means of a
transimpedance amplifier and a lockin amplifier. This method diminishes the effect of the
parasitic capacitances because $V_b$ is constant. A sketch of the circuit is shown in
Fig. \ref{circuit}. From the Lorentzian fit of the amplitude vs. frequency curve, $\nu_r$
and $Q$ are derived.

The resonant frequency depends on the springs constant $k$ and on
the system's moment of inertia $I$:
\begin{equation}
\nu_r = \frac{1}{2 \pi} \sqrt{\frac{k}{I}} .\label{fundamental}
\end{equation}

\begin{figure}[t]\begin{center}
\includegraphics[width=\linewidth]{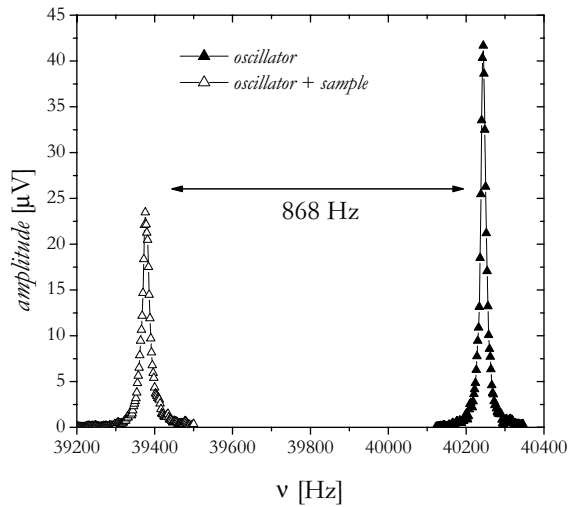}
\caption{Response of the oscillator with and without sample. When the sample is mounted
on top of the oscillator the system moment of inertia increases and the resonant
frequency decreases $868$ Hz.} \label{momentoinercia}
\end{center}\end{figure}

After the sample is glued to the oscillator, the system's moment of inertia is modified.
In our case, the sample is a single crystal BSCCO disk fabricated by lithographic
techniques and ion milling. The frequency variation was $\Delta \nu_r = 868$ Hz in
agreement with calculations (Fig. \ref{momentoinercia}).

On the other hand, $\nu_r$ is also modified by variations in the
effective $k$ of the system. If the sample is magnetic, an
external magnetic field $H$ produces an additional torque
\begin{equation}
\overrightarrow{\tau}=(\overrightarrow{M} \times
\overrightarrow{H})V = MHV \sin \theta,
\end{equation}
where $M$ is the magnetization, $V$ is the sample 's volume and
$\theta$ is the angle between $\overrightarrow{M}$ and
$\overrightarrow{H}$. For small $\theta$
\begin{equation}
\tau=MHV \theta =k_s \theta.
\end{equation}

The effective $k_e$ is obtained by adding $k_s$ to the spring
constant $k$. As a consequence, the resonant frequency increases
(decreases) if this torque is restoring (not restoring). The size,
geometry and anisotropy of the sample play an important role in
the magnitude and direction of its magnetization $M$. For example,
in a spherically symmetrical sample which is in the Meissner
state, $M$ remains antiparallel to $H$ during the whole period of
oscillation and the magnetic torque is zero ($k_s = 0$).
\begin{figure}[t]\begin{center}
\includegraphics[width=\linewidth]{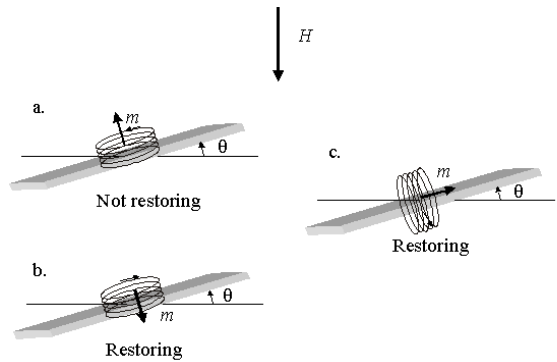}
\caption{Different behavior of the system when is applied an
external magnetic field. Currents in the layers of the sample are
depicted as a coil perpendicular to the plate and currents between
layers as a coil parallel to it} \label{coils}
\end{center}\end{figure}

A more complete analysis of the system response can be made by
supposing a constant current flowing in a coil placed on the
oscillator in the presence of $H$. The coil carries the original
current plus any current induced by the tilt of the oscillator and
generated by Lenz's Law. In this way, a change in the angle
between the coil and the magnetic field produces a change in the
induced current and therefore in the total current flowing in the
coil. In Fig. \ref{coils}.a and Fig. \ref{coils}.b we sketch the
case where $H$ and the coil are perpendicular to the plate. The
original current in the coil generates a magnetic moment $m$ that
interacts with $H$ exerting an additional torque, which is
restoring (not restoring) when $m$ and $H$ are parallel
(anti-parallel). The induced current in the coil produced by the
tilt of the oscillator is proportional to $1 - \cos \theta$. On
the other hand, when the coil is parallel to the plate and
perpendicular to $H$ as in Fig \ref{coils}.c the original $static$
current produces a change in the angle of equilibrium $\theta _0$.
However, the change in $k_e$ produced by this current is
negligible if $\theta _0$ is small as in this case. The current
induced by the alternating tilt of the oscillator produces a
restoring torque proportional to $\sin \theta$. Finally, when $H$
is parallel to the oscillator the torque is restoring (not
restoring) if $m$ and $H$ are parallel (anti-parallel). If $m$ and
$H$ are perpendicular the current induced by the alternating tilt
of the oscillator always produces a restoring torque.

The sample (BSCCO) has a layered structure. On mounting the disk,
the Cu-O layers ($ab$ planes) are parallel to the oscillator.
Following the previous description, currents in the layers can be
associated with a coil perpendicular to the oscillator and
currents between layers with a coil parallel to it.

\section{Results}
\label{}

\begin{figure}[t]\begin{center}
\includegraphics[width=\linewidth]{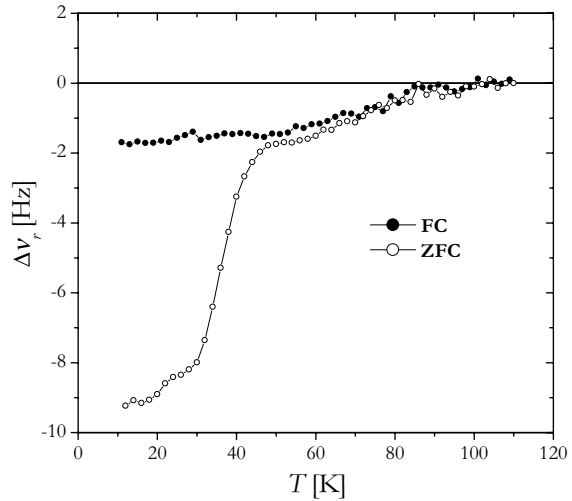}
\caption{Temperature dependence of the normalized resonant frequency for ZFC-FC
measurements with external magnetic field ($H=245$ Oe) perpendicular to the plate and to
the $ab$ planes of the sample. $\Delta \nu_r$ is always negative.} \label{perpendicular}
\end{center}\end{figure}

\begin{figure}[t]\begin{center}
\includegraphics[width=\linewidth]{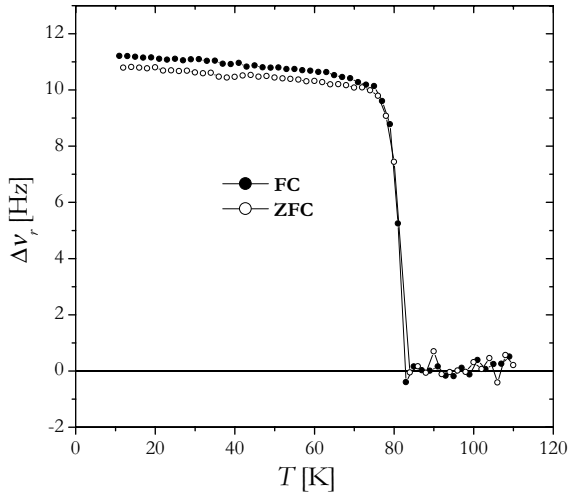}
\caption{Temperature dependence of the normalized resonant frequency for ZFC-FC
measurements with external magnetic field ($H=245$ Oe) parallel to the plate and to the
$ab$ planes of the sample. $\Delta \nu _r$ is always positive.} \label{parallel}
\end{center}\end{figure}

We did ZFC (zero field cooling) and FC (field cooling)
measurements. The difference $\Delta \nu _r$ between $\nu _r$
measured with and without an applied $H$ reflects the magnetic
response of the sample, eliminating any intrinsic effect in the
oscillator produced by changes in temperature. When an external
magnetic field is applied perpendicular to the $ab$ planes of the
sample, Meissner currents appear. These currents screen $H$
causing a non-restitutive torque (the intra layers current behaves
as in Fig. \ref{coils}.a). In this case, when the sample is in the
vortex state, $H$ penetrates the BSCCO forming pancake vortices
(PVs), and PVs in different layers are coupled weakly via
Josephson vortices (JVs). In Fig. \ref{perpendicular} we show the
measured $\Delta \nu_r vs. T$ data with $H=245$ Oe. $\Delta \nu
_r$ is always negative. When the temperature increases, the sample
becomes less diamagnetic and the magnitude of $\Delta \nu _r$
decreases. This is due to the entering of vortex in the sample. If
$H$ is parallel to the sample and to the oscillator $\Delta \nu
_r$ is always positive (Fig. \ref{parallel}). This is because the
thickness of the sample ($2.5$ microns) is smaller than the
penetration length in the $c$ direction ($30$ microns
approximately) and the magnetic moment of the JVs is negligible .
The only currents present in the sample are the Meissner ones,
that shield the alternating perpendicular $H$.
\section{Discussion and Conclusion}
\label{}

In the first case, when $H$ is perpendicular to the layers of the sample, $m$ is
proportional to $1 - \cos \theta$ (as we already said in the previous section). The
maximum deflection of the oscillator depends on the width of the plate and the minimum
distance between the plate and the electrode before {\em snap-down} takes place. This
phenomenon consist in the collapse of the plate with the bottom electrode and it happens
when the deflection reaches a third of the equilibrium distance between them
\cite{microsystem}. In our case the maximum possible $\theta$ is $\approx 1.5$ degrees
giving a maximum change in $m$ of $0.035$ $\%$. On the other hand, when $H$ is parallel
to the layers $m$ is induced by the tilt of the oscillator (perpendicularly to the planes
$ab$ of the sample) and is proportional to $\sin \theta$.

It is well-known \cite{Kes90} that in BSCCO $2D$ samples, the
phase transition to the vortex state depends on the component of
$H$ perpendicular to the $ab$ planes of the sample
($H_{c\parallel}$) and not on the component parallel to them.
Considering an error of $0.4$ Hz in the measurement of $\nu_r$ and
from the curves of Fig. \ref{perpendicular} we obtain $T_c (H_{c
\parallel}=245 Oe) = 78.7 \pm 0.7$ K. Due to the importance of
pinning, in the ZFC measurement $\nu_r$ is smaller than in the FC
measurement at low temperatures. When the temperature increases
these two curves become closer, and for $T > 46.4 \pm 0.7$ K the
behavior of the system is reversible. On the other hand, when $H$
is applied parallel to the sample (Fig. \ref{parallel}) the
measurements are almost reversible in all the range of
temperatures. In order to generate PVs, it is necessary to reach
the critical angle for which the perpendicular component of $H$
equals $H_{c1}$. The maximum component of $H$ perpendicular to the
$ab$ planes (appearing when the oscillator tilts) is of $6.5$ Oe
and, based on the observed reversibility, we can say that $H_{c1}$
is not reached.

These Si mechanical micro-oscillators are very sensitive magnetometers, suitable to study
mesoscopic high $T_c$ samples. The change in the resonant frequency ($\Delta \nu _r$) can
be positive or negative, depending on factors such as sample size, geometry and
anisotropy. These factors determine how the currents and magnetic moments are generated
in the sample.

\end{document}